\documentclass{sigchi}

\setcopyright{rightsretained}
\doi{}
\isbn{}
\usepackage{balance}       
\usepackage{graphics}      
\usepackage[T1]{fontenc}   
\usepackage{txfonts}
\usepackage{mathptmx}
\usepackage[pdflang={en-US},pdftex]{hyperref}
\usepackage{color}
\usepackage{booktabs}
\usepackage{textcomp}

\usepackage{microtype}        
\usepackage{ccicons}          
\usepackage[utf8]{inputenc} 

\usepackage[]{csquotes}
\renewcommand{\mkbegdispquote}[2]{\itshape}
\DeclareQuoteStyle[american]{english} {\itshape\textquotedblleft} [\textquotedblleft] {\textquotedblright} [0.05em] {\textquoteleft} {\textquoteright}

\usepackage{subcaption}

\usepackage{ccicons}

\usepackage{todonotes}

\def\plaintitle{Exploring Interactions with Voice-Controlled TV}

\def\emptyauthor{}
\def\plainkeywords{Authors' choice; of terms; separated; by
  semicolons; include commas, within terms only; required.}

\makeatletter
\def\url@leostyle{%
  \@ifundefined{selectfont}{
    \def\UrlFont{\sf}
  }{
    \def\UrlFont{\small\bf\ttfamily}
  }}
\makeatother
\def\pprw{8.5in}
\def\pprh{11in}

\definecolor{linkColor}{RGB}{6,125,233}
\hypersetup{%
  pdftitle={\plaintitle},
  pdfauthor={\emptyauthor},
  pdfkeywords={\plainkeywords},
  pdfdisplaydoctitle=true, 
  bookmarksnumbered,
  pdfstartview={FitH},
  colorlinks,
  citecolor=black,
  filecolor=black,
  linkcolor=black,
  urlcolor=black,
  breaklinks=true,
  hypertexnames=false
}

\begin{document}

\title{\plaintitle}

\numberofauthors{4}
\author{%
  \alignauthor{Sarah McRoberts\\
    \affaddr{University of Minnesota}\\
    \email{mcrob021@umn.edu}}\\
  \alignauthor{Joshua Wissbroecker \\
    \affaddr{University of Minnesota}\\
    \email{wissb004@umn.edu }}\\
  \alignauthor{Ruotong Wang\\
    \affaddr{Macalester College}\\
    \email{rwang2@macalester.edu}}\\
  \alignauthor{F. Maxwell Harper\\
    \affaddr{University of Minnesota}\\
    \email{max@umn.edu}}\\
}

\maketitle

\begin{abstract}
Intelligent agents such as Alexa, Siri, and Google Assistant are now built into streaming TV systems, allowing people to use voice input to navigate the increasingly complex set of apps available on a TV. However, these systems typically support a narrow range of control- and search-oriented commands, and do not support deeper recommendation or exploration queries. To learn about how people interact with a recommendation-oriented voice-controlled TV, we use research through design methods to explore an early prototype movie recommendation system where the only input modality is voice. We describe in-depth qualitative research sessions with 11 participants. We contribute implications for designers of voice-controlled TV: mitigating the drawbacks of voice-only interactions, navigating the tension between expressiveness and efficiency, and building voice-driven recommendation interfaces that facilitate exploration.

\end{abstract}

\section{Introduction}

With the emergence of streaming media applications like Netflix and YouTube, TVs (and the devices that power them) have gotten more capable and more complex. Instead of simply navigating a list of channels, users must navigate a set of apps, each with different organization and capabilities. Despite these changes, it remains uncommon to use a TV system with a touchscreen, keyboard, or mouse: the dominant paradigm is still to interact with TVs using a remote control, which has a limited range of inputs and has poor support for text entry and fine-grained navigation or selection.

Industry has developed several possible solutions to address the gap between interface complexity and the capabilities of remote controls. Among them is voice recognition technology: current systems from Apple, Amazon, and Google all support voice commands to change the channel and search for content. However, these voice user interfaces (VUIs) for TVs are optimized for command and control. We might imagine future interfaces that support more natural user interactions. For example, TVs could ask questions to clarify ambiguous inputs or elicit preferences, or they could adapt their verbosity or personality to the user's preferences. VUIs have an untapped potential to radically reshape the user experience of the modern TV interface.

However, there is little published research to help systems builders understand user acceptance and behavior around this topic. Prior work investigating natural language interfaces for TV has focused on systems issues \cite{fujita_new_2003} or application-specific questions \cite{renger_voistv:_2011, gordon_designing_2015}, rather than understanding user acceptance and behavior more broadly in the context of voice-controlled TVs.

To learn about user acceptance and expectations for voice-controlled TV, we developed a prototype system. The system is designed to mimic the experience of a modern TV application --- a movie recommender that helps the user find a movie to watch next. The system combines voice input with a hybrid voice and screen output. It is ``open ended'' in the range of inputs that it supports, to provide users a chance to phrase their needs as naturally or concisely as they wish.

We use a research through design approach \cite{zimmerman_research_2007} to learn from this prototype by conducting a series of 11 in-depth research sessions. Our methods are intentionally qualitative and exploratory, as we seek to gather rich experiential data to inspire future system designs and research studies. To organize our contributions, we explore the following research questions:

\begin{displayquote}
\textbf{RQ1:} How do users want to speak to a voice-controlled TV?
\end{displayquote}

\begin{displayquote}
\textbf{RQ2:} Do users prefer a voice-controlled TV to talk back or stay quiet?
\end{displayquote}

\begin{displayquote}
\textbf{RQ3:} How do users interact with a recommender system on a voice-controlled TV?
\end{displayquote}

In this work, we identify design lessons concerning voice input, a hybrid voice/screen output, and voice-driven recommender navigation. We conclude with design implications for VUIs and multimodal recommendation services. Our work highlights both the potential and roadblocks for more effective and engaging voice interactions.

\section{Related work}

Researchers have considered the potential of voice user interfaces (VUIs) from a human-computer interaction (HCI) perspective for decades. In this section, we first summarize related work in the domain of TV and recommender systems, then explore more broadly the research that informs our understanding of VUI design factors that impact the user experience.

\subsection{VUIs for TVs and Recommendations}

Voice user interfaces have been studied in many application domains (e.g., word processing~\cite{karl_speech_1993}, communication with robots~\cite{lee_receptionist_2010}, business meeting support~\cite{mcgregor_more_2017}, home automation~\cite{soronen_voice_2008}, mobile device accessibility~\cite{zhong_justspeak:_2014}, and question-answering~\cite{yarosh_children_2018}). Voice-controlled TV is one of these domains, with pioneering work in the early 2000s~\cite{fujita_new_2003} providing an early glimpse of the commercial devices --- such as the Apple TV and the Amazon Fire TV Cube --- that are available today. The application context of TV is interesting because users typically have access to an input device --- a remote control -- that provides only a coarse level of control and does not easily perform text input.

One very relevant study in the area of voice-controlled TV evaluates perceptions of a prototype system using a wizard-of-oz technique and a 2x2 design (voice vs. remote control; US vs. Japan)~\cite{tan_effects_2003}. This research finds that users generally evaluate voice interactions highly, but that there is an interaction effect where participants from the US prefer the VUI, while participants from Japan prefer the remote.

Other work in this domain has focused more on the development of innovative prototype systems (e.g., \cite{dalton_vote_2018, wissbroecker_early_2018}) and less on developing an understanding of user experience. For example, one paper describes a prototype system that allows users to post and read social media messages on a TV using voice~\cite{renger_voistv:_2011}, but this work exclusively focuses on implementation issues over HCI concerns. Another paper describes a prototype in-car system for helping families with children control a video player with voice~\cite{gordon_designing_2015}, finding with initial evaluation (N=1) that a child was able to use the system.

Increasingly, TV users interact with recommender systems, as apps from Netflix, Amazon, and other streaming services must help users navigate enormous media catalogues. A dominant thread in prior work that combines natural language interfaces with recommender systems concerns ``conversational'' recommenders~\cite{mcginty_adaptive_2006} that allow the user to critique search results. Some early systems apply this technique to recommend restaurants~\cite{goker_personalized_2000, thompson_personalized_2004}, finding that a personalized system was able to reduce the number of voice interactions required to find a place to eat, as compared with a non-personalized control condition. Another early study developed a prototype text-based recommender on an e-commerce site, finding that users are able to issue relatively complicated textual critiques, such as relational critiques~\cite{chai_natural_2002}. Recent work has investigated the use of modern machine learning techniques on user review data to support more coherent and flexible conversational recommendation dialogues~\cite{reschke_generating_2013, zhang_towards_2018}.

There are only a few examples of user-centric research in the space of natural language recommendation systems. One paper developed a hierarchy of natural language requests by collecting a dataset of queries from users, categorizing query features as objective, subjective, or navigation~\cite{kang_understanding_2017}. Other work looked to online discussion forums to better understand the nature of ``narrative-driven'' recommendation requests~\cite{bogers_defining_2017}, examining the frequency of common patterns.

\subsubsection{Contributions.}
Prior work on VUIs in the context of TV and recommendations have primarily emphasized the development of innovative prototypes over learning from user studies. Therefore, there are still large open questions concerning how to design for these types of applications. In this work, we conduct a qualitative user study to answer questions about user acceptance broadly, and to investigate two specific unanswered questions.  First, how do users interact with a movie recommender system that lacks any input modality except for voice? Second, do users prefer a such an application to talk back, or to remain silent? 

\subsection{User Studies of VUIs}

Because of technical complexities, it remains difficult to build high-quality natural language interfaces; HCI researchers have taken several approaches to conducting studies in this environment. Some prior work has employed research through design methods~\cite{zimmerman_research_2007} to prototype new interfaces and learn about the resulting user behaviors or technology acceptance. In some cases, researchers have used Wizard of Oz methods~\cite{tan_effects_2003, yarosh_children_2018, vtyurina_exploring_2018} or technical probes~\cite{mcgregor_more_2017} to accelerate the introduction of novel technologies into a context suited for HCI research. Other studies were built on full research prototypes, such as a receptionist robot~\cite{gockley_designing_2005} and a mobile application built to provide accessibility through voice~\cite{corbett_what_2016}.

Prior work has looked broadly at user acceptance of VUIs. Several recent studies have examined the use of VUIs in homes to better understand interaction patterns and design implications. One paper describes a wide set of quantitative and interview results concerning the use of Amazon Echo devices~\cite{sciuto_hey_2018}, providing insights into different ways in which the devices are incorporated into daily life. Another recent paper looks at a similar topic, focusing on how families interact together --- often unsuccessfully --- with their device~\cite{porcheron_voice_2018}, arguing that these interactions are not ``conversational,'' but rather ``request/response'' in nature~\cite{reeves_this_2018}.

It is a common and unsolved design challenge to help users discover and learn new commands in a VUI~\cite{pradhan_accessibility_2018}. Users may assume that the system supports a broader set of commands than it actually does, or users may not know how to access some types of functionality~\cite{yankelovich_how_1996}. One possible direction, explored in the context of accessing mobile applications through voice, is to provide graphical overlays with visual prompts for the user~\cite{corbett_what_2016}. In another study, researchers used a portion of a graphical display to show a ``what can I do'' menu to assist users in discovering commands~\cite{furqan_learnability_2017}.

Another set of open design questions concern the incorporation of personality into a VUI. There is evidence that many users personify intelligent assistants like Alexa in the Amazon Echo~\cite{lopatovska_personification_2018, purington_alexa_2017}. One recent study investigated differences in attitudes between children and adults, finding that children have a stronger preference for personification~\cite{yarosh_children_2018}. One specific aspect of personification is the support~\cite{vincent_google_2018} or encouragement~\cite{bonfert_if_2018} for polite interactions.

Finally, it is an open design challenge to support users when the VUI fails due to incorrect recognition, an unsupported command, or other factors. One recent paper explored this issue in the context of music search, a domain with many artists and song titles that are difficult or impossible to say unambiguously~\cite{springer_play_2018}; they describe methods for automatically identifying content that is not discoverable through voice, and for adding voice support through crowdsourced pronunciation. Another recent paper documents different types of errors, and different tactics that users employ to recover from error states~\cite{myers_patterns_2018}.

\subsubsection{Contributions.}
Much recent research has been dedicated to understanding the user experience around VUIs, describing a variety of problems in current systems, and exploring different design decisions. In this research, we use now-established HCI methods (i.e., prototyping and in-depth qualitative methods) to add new findings that speak to several of these themes. In particular, we investigate participants' impressions of (and desire for) designs around command discoverability, the incorporation of personality, and error recovery.

\section{Research Platform}
\label{research-platform}

\begin{figure}[t]
  \centering

  \begin{subfigure}[b]{\columnwidth}
    \includegraphics[width=\columnwidth]{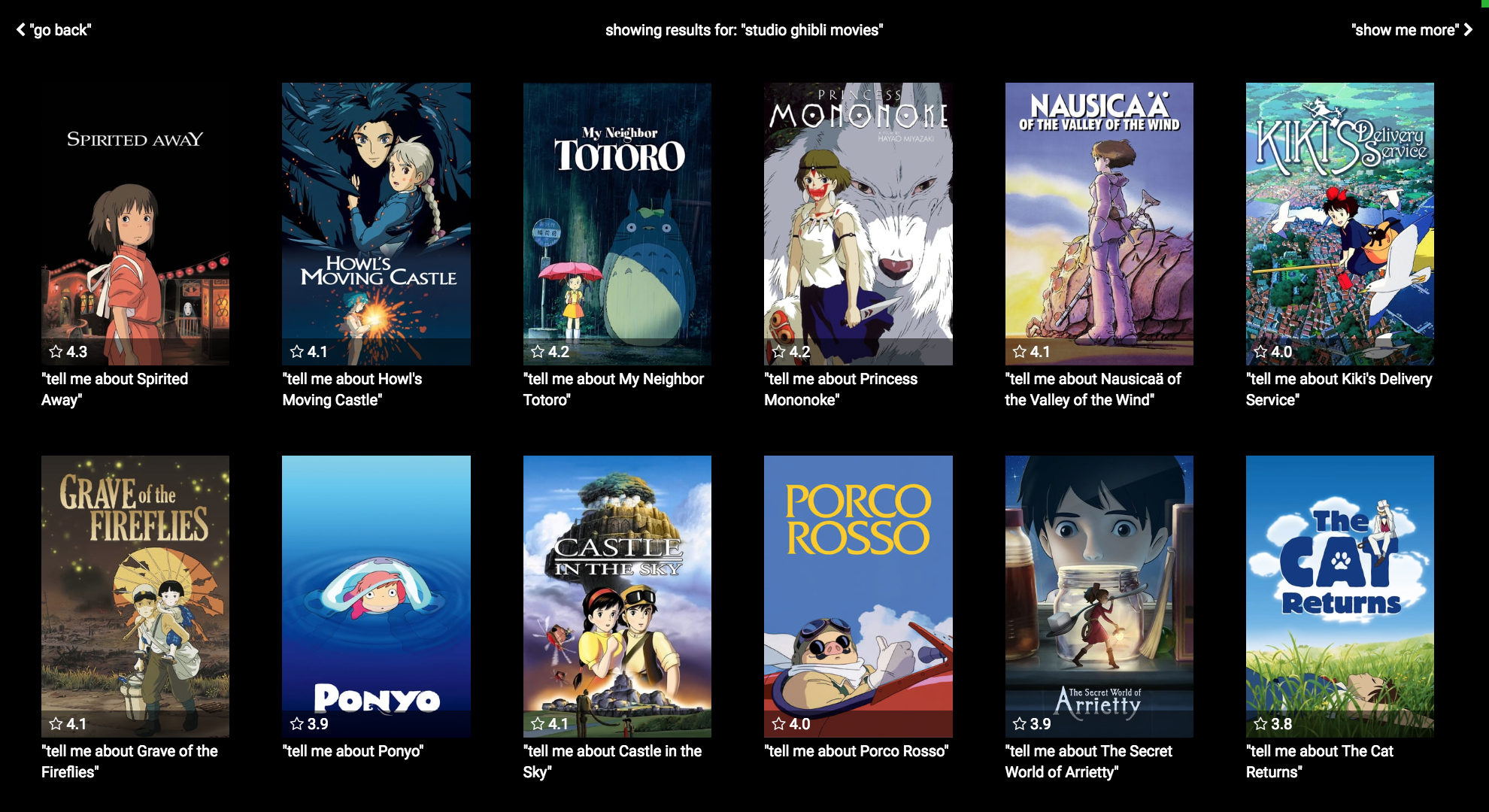}
    \label{fig:screenshots-top}
  \end{subfigure}

  \begin{subfigure}[b]{\columnwidth}
    \includegraphics[width=\columnwidth]{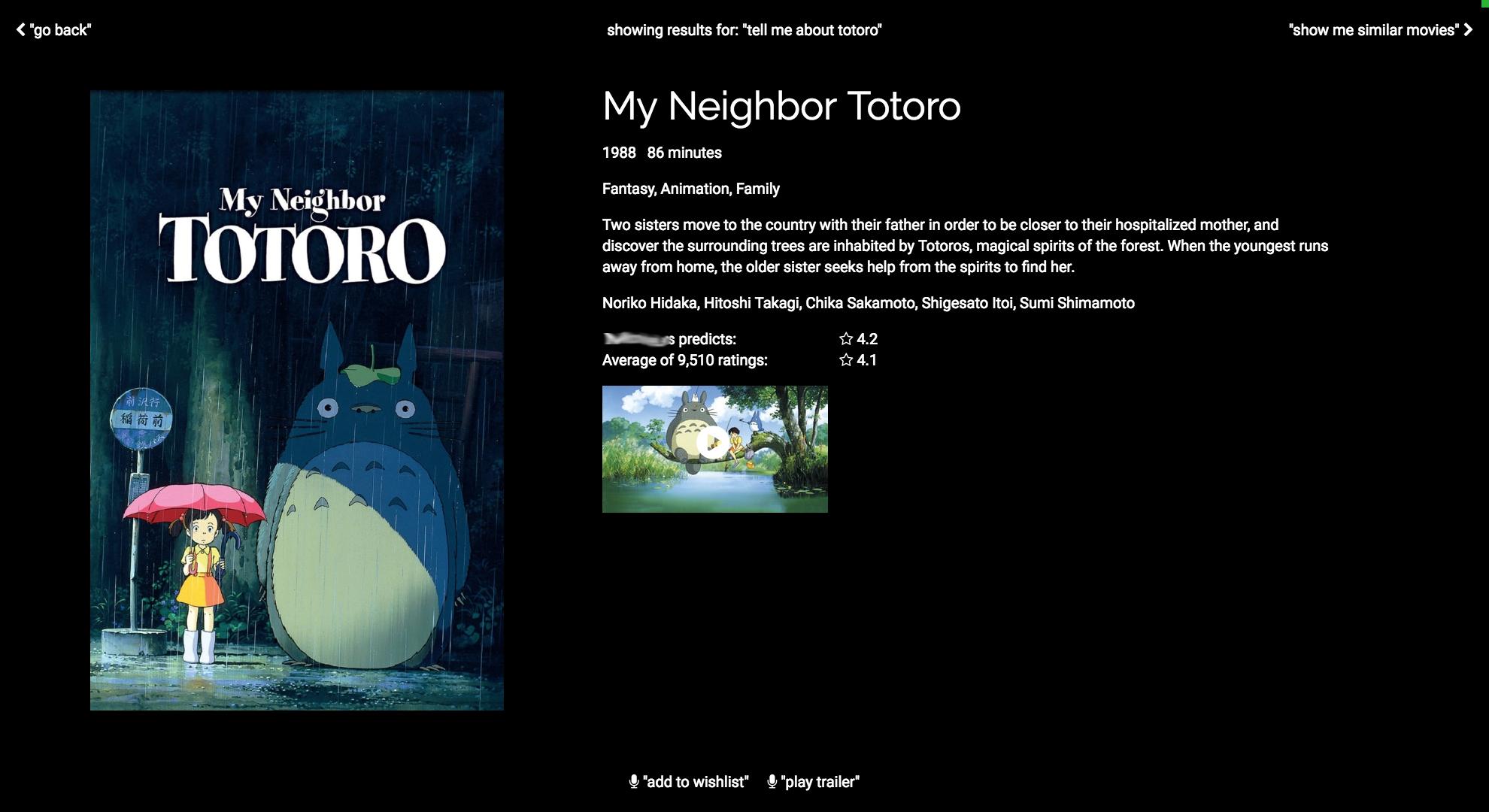}
  \end{subfigure}

  \caption{Screenshots of the MovieLens TV interfaces: the explore view (top) and the details view (bottom). Because there are no available input modalities except for voice, we label available actions in the interface in quotes (e.g., ``go back'').}
  \label{fig:screenshots}
\end{figure}

To learn about user acceptance of voice-controlled TV, we built a prototype system called MovieLens TV using web technologies and an Amazon Echo --- see Figure~\ref{fig:screenshots} for screenshots. MovieLens TV is designed to help its users find a movie to watch by combining a voice user interface with an on-screen interface to visualize results and play movie trailers. There are no input modalities available to users except voice --- e.g., to navigate to the previous page the user can say ``go back''. MovieLens TV is optimized for display on a large widescreen television, with a black background, large images and text, and no scrolling.

MovieLens TV is built on top of a custom Amazon Alexa skill that allows users to ask an Amazon Echo smart speaker for movie recommendations. Because custom Alexa skills have access to only a very limited user interface, we instead use our own web interface. To integrate the web interface and the smart speaker, our server that handles communication with Alexa also maintains WebSocket connections with active browsers: simultaneous to pushing voice responses to Alexa, it pushes WebSocket messages to connected clients. The JavaScript client software changes the user interface in response to these messages. A user sees the user interface change simultaneous with hearing a voice response from the smart speaker. More details about the system architecture are available in~\cite{wissbroecker_early_2018}.

\begin{figure}[t]
  \centering
  \includegraphics[width=\columnwidth]{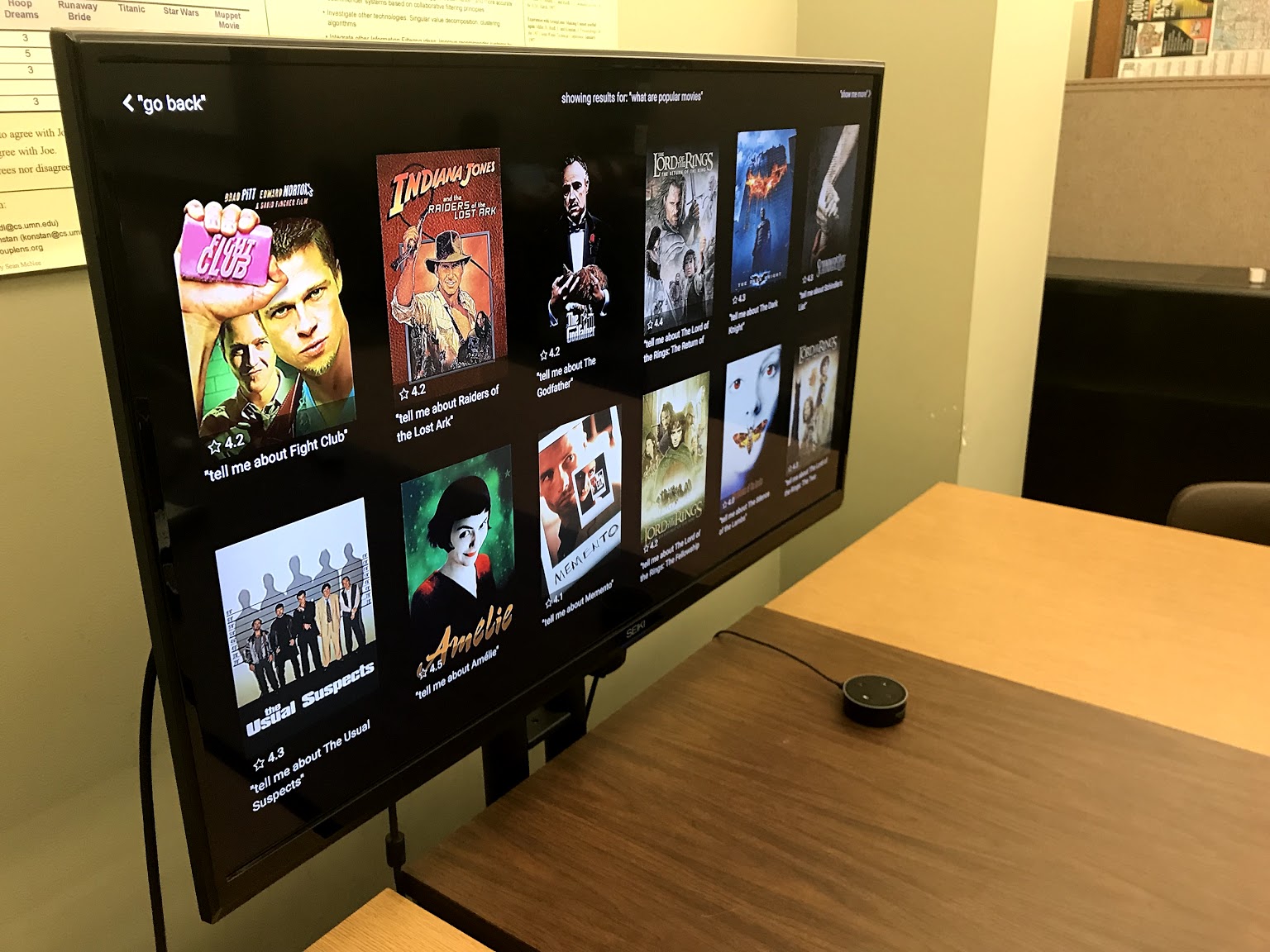}
  \caption{A deployment of MovieLens TV for user testing. The voice input and output device, an Amazon Echo Dot, is shown on the table.}
  \label{fig:tv}
\end{figure}

Using Amazon's off-the-shelf hardware allows us to prototype a TV with voice recognition built-in without actually having such a TV, and allows us to leverage Amazon's excellent microphones and speech-to-text transcription service. To perform user testing, we require a TV (running a web browser in full-screen mode) and an Amazon Echo Dot, shown in Figure~\ref{fig:tv}.

To use the system, a user must first install our custom Alexa skill and link it to their MovieLens account using standard practices. MovieLens is a personalized movie recommendation site with thousands of active monthly users. Once the Amazon and MovieLens accounts are linked, responses to user requests may be personalized according to their recommendation profile (i.e., different users issuing the same query will usually get different results). After this one-time setup, the user must say ``Alexa, open MovieLens'' to launch the voice interface. At this point, the user may begin sending requests, prefixed by ``Alexa'', to ensure that the Echo is listening. For example, the user might say ``Alexa, show action movies'', followed by ``Alexa, show me more''. The Echo's ``follow-up mode'' feature~\cite{crist_alexas_2018} was released during our development and testing, so in our evaluation we required each verbal request to begin with the Alexa wake-word.

MovieLens TV has several views (see Figure~\ref{fig:screenshots} for two examples), including an explore view for displaying movies in a grid, a details view for viewing information about one movie, a trailer view for watching movie preview videos, and a home/help screen. Simultaneous with showing the view, the Echo (optionally) vocalizes a response. For example, in response to the query ``show me futuristic movies'', the Echo says ``here are some movies that I think are futuristic''. The application maintains user history, allowing users to say ``Alexa, go back'' to visit previous pages whenever they wish.

Because new users will not understand which commands will work well, we include suggestions in the user interface. To help users get started, the home page contains three instructions: 
(1) \textit{Ask for movie searches starting with ``Alexa''. For example: ``Alexa, show me action movies''}, (2) \textit{Say ``help'' any time to learn more about what you can do.}, (3) \textit{Each screen will show you things you can say in ``quotes''. Try these out!}


As indicated in these instructions, the UI indicates some of the available voice commands explicitly, as is recommended in early design research~\cite{yankelovich_how_1996}. We visually emphasize available commands using quotation marks, in some cases adding a microphone icon for emphasis. For instance, different views suggest actions such as ``go back'', ``see more'', ``play trailer'', and ``show similar movies''. To help users navigate to a particular movie, each movie image is accompanied by a command, such as ``tell me about Avatar''.

Users are free to say whatever they wish to the system. To infer meaning from what users say, we use \href{https://wit.ai}{wit.ai}, a third party intent detection service that uses machine learning to infer the ``intent'' of each request based on our small library of internally developed training examples. For instance, we map ``show me Toy Story'' to an intent to view details of a particular movie, while we map ``show me movies like Toy Story'' to an intent to view related-item recommendations. Wit.ai also detects a variety of entities in each query, such as references to genres, actors, movie titles, time frames, or other filtering and sorting criteria. Our server accepts the wit.ai structured representation of the user's query, and converts it into a decision about (a) which view to show, and (b) what content to show in that view. The mapping between intents and view logic is simply a set of if/else rules. For searches, unless we detect an explicit sort order (e.g., ``recent'', or ``popular''), we rank results using the MovieLens item-based K-nearest neighbors collaborative filtering algorithm, a classical recommendation technique~\cite{sarwar_item-based_2001}.

\section{Methods}

Our goal in developing the prototype of MovieLens TV is to better understand how users interact with a voice-controlled TV. This is a form of design research~\cite{zimmerman_research_2007}, intended to be qualitative and exploratory: the prototype is imperfect, and we did not know going into the study the degree of its usability or utility. Therefore, to collect data, we conducted 11 interview and observation sessions, where we talked with participants broadly about their online movie-finding habits, and asked them to interact with the system.

\subsection{Interview and Observation Sessions}

We used convenience sampling --- online flyers and in-person recruiting in computer science classes --- to recruit 11 undergraduate participants (5 female, 6 male) at a midwest university. All participants reported access to intelligent assistants through their smartphone (e.g., Siri or Google Assistant), but only two had access to an intelligent assistant in a smart speaker (e.g., Amazon Echo or Google Home). Two participants reported being frequent users of intelligent assistants, while seven reported being occasional users, and two had only minimal experience. All users reported familiarity with finding TV and movies to watch using an online service (e.g, Netflix or Hulu), including seven regular users and four occasional users.

We conducted the interview and observation sessions in a maker space. We ran the prototype using a 42 inch flat screen TV connected to a laptop computer (displaying a Google Chrome browser window in full-screen mode), and an Amazon Echo Dot. We used a separate microphone to record the interviews. 

Each session began with a pre-interview, and ended with a post-interview. The pre-interview (\textasciitilde10 minutes) focused on participants' general movie finding strategies and prior experience with voice assistants and movie recommender systems, while the post-interview (\textasciitilde10 minutes) asked participants to reflect on their experience using MovieLens TV.

The main portion of each session (\textasciitilde40 minutes) asked participants to complete two movie finding tasks, using MovieLens TV. To begin this procedure, we briefly helped the participant to create a personalized movie preference profile using the existing new user setup interface in MovieLens. We then asked the participant to find a movie to watch, using MovieLens TV however they wanted. We asked users to talk through their actions. The interviewer asked questions of the participant throughout, especially at times when the participant was thinking or appeared confused by how to proceed. The participant would complete each task by telling the interviewer that he or she had found a movie to watch.

We asked participants to perform this task twice in order to manipulate one aspect of the interface: whether or not the system ``talks back''. The two conditions are:

\begin{itemize}  
\item \textbf{visual+voice}: the Echo vocalizes responses to queries (e.g., ``here are some futuristic movies''), in addition to the visual user interface on the TV.
\item \textbf{visual-only}: the Echo remains silent throughout, leaving only the visual user interface to display results.
\end{itemize}

We exposed each participant to both conditions. We chose the order randomly --- 5 participants saw visual+voice first, while the other 6 saw visual-only first.

We transcribed the study sessions using \href{https://sonix.ai}{sonix.ai}, a partially automatic transcription tool. The lead author read and open coded~\cite{muller_survey_2014} the transcripts, resulting in 589 codes from the 11 interviews. Three authors then analyzed these open codes, and then applied a constant comparison affinity mapping approach~\cite{dye_constant_2000} to identify recurring themes. In addition, we extracted the text of participants' queries to the Amazon Echo device to perform descriptive, quantitative analysis of the frequencies of different types of queries.

\subsection{Limitations}

This is an exploratory, qualitative investigation of an early prototype, with the goal of developing a set of themes that can motivate and inform more detailed future work. Our chosen methods have several inherent limitations, including a limited number of participants and a short window for observation.

We have chosen a single application domain --- movie recommendations --- which limits the scope of some of our results. While challenges with speech recognition on entities such as names may generalize to other domains (e.g., streaming music~\cite{springer_play_2018}), other features are domain-specific (e.g., academy award winners). Part of our goal is to learn about voice user interfaces on a television, and the domain space of recommendations biases our findings towards a particular type of use (e.g., searching for content in a single domain) over other types (e.g., issuing commands to turn the system on and off, or to switch input sources).

\section{Results}

To address the three research questions, we synthesize how participants used and responded to the MovieLens TV system. In this section, we report themes, participant quotes, and quantitative results about the queries spoken to the TV.

\subsection{RQ1: How Users Want to Speak to the TV}
The only input modality supported by MovieLens TV is voice. This novel interaction design allows us to learn how participants initially react to the interface, and to learn how they adapt their behavior as they learn the interface and encounter successes or failures with particular spoken commands.

\subsubsection{Query Counts}
The results below capture specific examples from a much broader array of commands. To provide some context for broader application usage, we tally how often participants voiced different types of queries. Across the sessions, 11 participants issued 504 commands. The most common commands involved asking about a particular movie (11/11 participants, N=116), searching for a genre or category of movie (10/11, N=59), searching for related or similar movies (10/11, N=54), and searching by actor or director (7/11, N=30). Participants also frequently issued navigation commands, including variations on ``go back'' (11/11, N=80) and ``show more'' (11/11, N=70). Commands tended to be short, with a median of four words (excluding the wake word ``Alexa'').



\subsubsection{Discovering Commands}
The interface included hints about available commands (see Section~\ref{research-platform} above) throughout (e.g., ``tell me about [movie\_title]'' under movies and ``$\leftarrow$ go back'' in the top left). Eight participants expressed their appreciation for these hints. P6 had ideas for more ways to suggest commands, suggesting the inclusion of ``show me more movies with Tom Hanks'' on the information page for a Tom Hanks movie. Seven participants asked for more instructions in a help page or manual. P6 wished there were more command suggestions at the beginning. P3 wished it was possible to say ``show me commands'' to discover more in situ recommendations~\cite{corbett_what_2016}, describing, \textquote{have something up there like a little quote like where it says `say show me more' you could just have `show me commands' and then a drop list would come down.}

While the command hints were successful at helping participants navigate the interface, the open-ended support for search commands actually created some feelings of uncertainty or anxiety. As P2 said, \textquote{I don’t want to [ask it to] do things it won’t do.} Other participants felt lost learning commands. P8 reflected on learning and trying new commands, \textquote{I didn't know the commands. Some of the commands I would have thought would be obvious weren't that obvious.} Some participants didn't know what they wanted to say when they started talking. P3 started, then quickly gave up on a command, saying, \textquote{Alexa \ldots I don’t know}.

Another outcome of open-ended query support was that participants issued a wide variety of queries that MovieLens TV did not support, either because we had not yet encountered that particular phrasing in our earlier internal testing, or because it is a system feature that we had not built. For example, the following queries were not supported at the time of the study session: ``Alexa, get out of the video'', ``Alexa, what is the source of the ratings?'', ``Alexa, read description'', and ``Alexa, show me the ratings''.

\subsubsection{Limitations of Voice Input}
There are several drawbacks to relying on voice as the sole input modality. Our research sessions uncovered several themes related to these drawbacks: difficulties resolving failed commands, irritations with cumbersome language, and desires for more efficient interactions.

Perhaps the most conspicuous source of VUI problems concerns failure. In some cases, the failure was due to incorrect speech recognition, while in other cases, it was caused by an unsupported or misinterpreted query. These errors disrupted participants' task flow. Participants attempted to recover from failures in different ways. One common action was to simply try the same query again (10/11, N=60). However, repeating the query worsened the experience, as P7 explained: \textquote{It’s annoying to have to repeat yourself multiple times, especially when you think you’re sounding clear.} Other participants adjusted their query due to the failure. For instance, P8 asked \textquote{Alexa, find Sherlock Holmes}, followed by \textquote{Alexa, show me the series of Sherlock Holmes}, to locate a particular version of that title. P1 started their session by asking \textquote{Alexa, I'm ready to find a movie}, then trying \textquote{Alexa, show me some movies} when that command was not correctly recognized. These errors jarred the experience of voice-input and detracted from the quality of the experience. P1 didn't notice the drawbacks of voice input until things went wrong saying, \textquote{That was the first time it hit me that I was using a different [input] method. It was mindless up to that point.} 

Cumbersome or repetitive commands also worsened the user experience. Particularly, commands like ``go back'', ``show me more'', and ``tell me about'' felt wasteful to participants. P11, when searching for the newest Spiderman movie, had to repeatedly ask the system for the ``next one'' in order to pass through six other Spiderman movies that they didn't want. Additionally, three participants complained about the ``go back'' function, especially when they issued the command more than once consecutively. P7 explained that asking the system to go back is similar to but more work than going back on a webpage, \textquote{it's kind of annoying. It's like the back on a computer. You know, all the time I want to go back four pages, and I say go back go back.} After watching a trailer, P5 expected the system to automatically begin the next activity (like starting the movie, or returning to the information page), \textquote{Go back. This is kind of different that you have to say `go back' after watching the trailer.} Additionally, since our version of Amazon Echo required the frequent use of the ``Alexa'' wake word, this lead to more laborious interactions. As P4 said, \textquote{It wastes time saying its name}. 

Participants talked about several possible solutions to these limitations. Four participants who had trouble with the system understanding them wished that they could click on or type the option they want. While three participants wanted to say other context clues on the screen to help specify a movie if there was any confusion (E.g. ``movie number 4'' or ``the one with Matt Damon'' while looking at a list of movies). To streamline cumbersome interactions, two participants started thinking of their own shorter commands. P8 wanted to be able to cut words out of their commands, saying \textquote{If I say `The Matrix' if it's on the page, it should directly go to The Matrix. I shouldn't be saying `tell me about The Matrix'.} P11 got bogged down by ``show me more'' and wanted to be able to say ``next'' instead. Ultimately, P11 wanted a shorter way to say commands and complete interactions, saying, \textquote{I think shortening how to say things could be potentially beneficial.}

\subsection{RQ2: Talk Back or Stay Quiet?}

As described above in Methods, we asked each participant to use MovieLens TV in two conditions: one where Alexa vocalizes responses (``visual+voice''), and one with no audio output (``visual-only''). Preferences between these two conditions were split across participants (5 preferred visual-only, 4 preferred visual+voice, 2 were undecided). As we asked participants about their experience, we learned that what a system says is more important than whether it says anything.

\subsubsection{Benefits of Voice Output}
Some participants found the voice output to be a positive addition to the experience. P5 felt a little less silly to be talking to something while it’s talking back, \textquote{[in the visual+voice condition] there was a lot more confirmation that helped out with finding when there were similar titles, that sort of thing. I don't know, it felt almost less silly, instead of me just talking to something like `oh it's kind of talking back a little bit.'}

Another benefit of voice output is adding clarity to the source of misunderstandings, including when the system needed more help understanding what movie participants were looking for. P7 couldn't tell what to do, on the silent first version, when there were multiple matches for the movie ``Life'' but had a much better understanding when the system prompted the issue out loud later. P2 also experienced this contrast, \textquote{I liked when I would say `Tell me about this movie,’ it would say if there was more than one option, because it didn't really put up anything on the screen saying that was it or if there were more.}

Participants also wanted to have the option to listen to the app instead of read content. Five participants thought it would be helpful if the system could read the description or talk more if they weren't able to look at the screen, or didn't want to read the description. P8 explained, \textquote{if I'm cooking [or] eating food or just taking a rest, just read it to me.} P10 wanted the system to only speak when requested, explaining, \textquote{That was more me. I guess that will be the only time that I would say [I] want to hear it back.}

\subsubsection{When Voice Output Is Unnecessary}
Many of the spoken responses of the system felt useless to participants. Reflecting on her experience, P10 remembered tuning out the system when it was talking, \textquote{Now that you bring it up, I remember it saying stuff. Yeah. But [\ldots] it makes me kind of feel bad, because I just don't pay attention when it was an automated thing and it's talking, so I didn't really notice.} However, not all participants were able to so naturally ignore the visual+voice condition. Six participants complained about when Alexa said redundant information. Phrases that people found most annoying were statements presenting what's already on the screen, like ``Here is The Martian.'' \textquote{I'm like `yeah, I know it's up there,}' said P7. P1 demonstrated this further with a search through action movies: \begin{displayquote}[P1]
Ok. Alexa, show me action movies. [Alexa: here are some movies that I think are action] Definitely like no voice. And I said show me more and she just tells me here is some more action. Alexa, show me more. [Alexa: here are some more movies that I think are action] Alexa, show me more. [Alexa: here are some more movies that I think are action] [\ldots] `Here are some movies that I think are action' is especially repetitive if users are just loading the next page.\end{displayquote}  The drawbacks of voice output are highlighted by visuals in the system. Voice output is not essential in a system with visuals, so making sure that what’s said enhances the system is important to the participants.

The importance of efficiency also impacted participants' opinion of voice output. \textquote{I don’t like to waste my time,} said P8 about the interactions. Four participants cautioned that efficiently finding a movie was more important than experiencing the VUI. P7 also explained \textquote{For most things, just don't say anything at all. Yeah. I feel like that's the best way to go. And then if it needs more information I can ask you anything. [\ldots] It doesn't need to be like a person.} For these participants, voice outputs should not detract from the task at hand.

\subsubsection{Conversational Aspects and Personality}
Seven participants showed interest in some level of personality in their interactions. For example, P7 thought it would be funny if the system had some opinionated answers built in, to support actions like asking Alexa if she has a favorite movie. P7 elaborated further, \textquote{That's just kind of a thing where it could give a random movie. That would be kind of fun. [\ldots] Almost give it opinions. [\ldots] So have it be biased towards certain genres.} During the study, P3 was already projecting personality on the system, explaining \textquote{Yeah totally, that’s kinda what I was poking at like `Alexa, I don't know’ and Alexa was like `make up your mind!’ Things like that are kinda fun honestly. It would just get obnoxious if that chimed in while you were trying to find a movie, but yeah, that would totally be awesome.} P1 even thought that a personality could help make the voice output more worthwhile saying, \textquote{If it had a personality then I guess it would be better with the voice. [\ldots] A couple sentences. laid back. I guess that'd be cool.}

Other participants were more concerned with polite interactions and a personable experience. P2 reflected on wanting to be able to say ``please'' and ``thank you'' to the system. \textquote{I feel like it's being rude!. I know it's not real\ldots Like I want to say all the pleases and stuff without it getting confused.} P6 specifically wanted the system to be more personal and personable (e.g., including her name) perhaps saying ``I have another movie you might like.'' But only to an extent; everybody mentioned that sometimes it would be inappropriate and get in the way to have the system respond with personality or interject opinions. Still, supporting prepared answers to joke questions could be a lightweight way to include personality, fun, and personable features without getting in the way of tasks.

\subsection{RQ3: How Users Interact with the Recommender}
Unlike a typical TV-based recommender (e.g., Netflix or Amazon Prime Video), our prototype system only supports voice input. In this section, we report on how subjects requested and evaluated recommendations using voice.
\subsubsection{Patterns of Use}

Participants wanted the system to provide more ideas about possible searches. In particular, six participants shared that the home screen would have been improved by showing content (e.g., top recommendations), to give them ideas about the types of content they were interested in finding. As P4 stated, \textquote{At the very beginning I don’t know what kind of movie I wanted to watch.} This home page of recommendations would also serve as a reset if a user doesn't feel directed in their movie search:\begin{displayquote}[P5]
I wouldn't always [\ldots] know what to search for specifically.\end{displayquote}

Once participants got started, we observed two recurring browsing patterns. First, participants issue queries in a series, following a promising lead or information scent~\cite{pirolli_information_1999, chi_using_2001}. For example, P3 started from an actor (``Alexa, show me Jesse Eisenberg''), then looked at one of the search results (``Alexa, tell me about the Social Network''), then moved to a different actor search (``Alexa, show me movies with Justin Timberlake''). Another pattern involves executing a search, then examining multiple items in turn. For example, P5 searched by genre (``Alexa, show me musicals''), then proceeded to ask about several movies in turn (Sweeney Todd, Moulin Rouge, then Chicago, each followed by a ``go back'' request) before finally moving to a different search (while viewing the movie Chicago: ``Alexa, show me more like this'').

\subsubsection{Balancing Familiarity with Novelty}

We noticed that participants evaluated their first searches based on how many movies came up that they already knew and liked. As we asked the participants to talk aloud while they looked at their search results, we heard reactions similar to this one for movies with Leonardo DeCaprio:\begin{displayquote}[P9]
Ok. I actually like\ldots I like Wolf of Wall Street. I like Django Unchained. Catch Me If You Can. I want to see Inception. I wanted to see The Revenant. I wanted to see The Great Gatsby but that's probably on another page. [\ldots] Yeah, I like a lot of them. Never heard of The Aviator actually\ldots\end{displayquote}
Talking about the recommendations after the fact, P5 reflected, \textquote{It was bringing up stuff I'd already wanted to see [\ldots] that was really nice, and really well done.} For some participants, seeing familiar movies made their selection easy. P6 wanted to watch a movie that was familiar, and explained the serendipity of the system’s recommendation, \textquote{Just because it's like something that I've been wanting to watch for a while. And now there it is, so it's like oh yeah. Perfect. There we go.}

Not only did participants evaluate the recommendations according to familiarity, but some also used familiar movies to inspire exploration. P2 took time to explore ``tell me about'' and ``show similar movies'' for a collection of spy movies she had seen before. Despite the familiarity, she was impressed by the accuracy of the recommendations. P2 also tried watching the trailer for an old movie she already knew. What's more, a group of familiar movies occasionally served more as a motivation to keep digging, to uncover new movies. P9 continued a search through action movies after seeing a familiar franchise, \textquote{Alexa, show me action movies. Ok, I do like Star Wars, I have so much Star Wars stuff. Alexa, show me more.} We found that seeing one new movie in a group of familiar ones can pique a user's interest. As P7 reacted to a collection of movie recommendations about ``space,'' the unknown movie in the set of 12 was the most eye-catching. As P7 described, \textquote{Let's see. There are good recommendations. Alexa, tell me about Moon\ldots It's the only one here I haven't heard about.}

Participants looking for something new would leverage their movie-watching experience by searching for movies similar to their favorites. P6 shared, \textquote{I would say, [recommendations] that are more interesting are the ones that are similar to the first movie, the Guardians of the Galaxy.}  However, we found that too much familiarity could hamper the user experience. P11 happened to have already seen many of the movies that came up in searches, so it was harder for this participant to find new movies to watch. Many movie recommender systems do not show already-rated or watched content, and this participant agreed that this would be useful: \begin{displayquote}[P11]
I'd seen the majority of the superhero movies on there. I don't want to watch them again. I think it'd be nice to be able to filter them out as a whole.\end{displayquote}

\section{Discussion}
\subsection{Implications for Efficient Interactions}
Participants found several voice input patterns to be inefficient. Some navigation commands (e.g., ``go back'') were slow to respond, and sometimes required saying the same phrase over and over. Content selection commands (e.g., ``tell me about [movie title]'') required too many words, and were prone to speech-to-text errors. 
We think systems should leverage mixed-input approaches to combine the strengths of voice input (e.g., expressive searches) with the strengths of physical input devices (e.g., push to talk, content selection, and interface navigation). It remains future work remains to develop voice-only experiences that are more responsive and efficient.

We compared a ``visual+voice'' system with a ``visual-only'' system to learn about how people perceive a TV talking back to them. We found that participants had a mix of reactions, some appreciating the voice output, and others finding it annoying. One implication of our study is that simple responses such as ``here are some movies that I think are comedy'' have little value, and should be replaced by quicker audio feedback, such as a chime. Several participants felt, however, that voice outputs helped to emphasize key data, such as the fact that the system had found multiple matches for a search, or that a movie had an especially high predicted rating. Developing a better understanding of how selective voice output can improve the user experience around visual user interfaces is interesting future work.

\subsection{Implications for Expressive Interactions}

Although efficient interactions were important to some users, others showed interest in more expressive and personable interactions. The visual+voice condition served an important role in clarifying miscommunication, and its conversational aspects were engaging and comfortable for some users. Participants used joke questions and polite language to demonstrate their interest in the more expressive side of voice interaction.
 
As some participants explained, even obvious voice output is sometimes still worth hearing. If a user is too far from the TV to read the screen, or is looking down at food, it could be helpful to have the system describe the content, or read descriptions. While it is out of the scope of this work, it is also important to consider how visual and audio cues work together to suit accessibility needs~\cite{rector_eyes-free_2017}. 

Since there is a tension between the design of efficient and expressive interactions, we think it is important to understand users' context and preferences. One simple way to address this tension is to allow users to adjust a verbosity or personality setting, a possible feature that several participants brought up. Alternatively, it is interesting future work to develop adaptive systems that can understand short-term cues or long-term preferences indicating a desire for expressive interactions (joke questions, polite language), or efficient interactions (terse commands). 

\subsection{Implications for Recommender Systems}
One of the most successful elements of the prototype, according to participants, was its recommender system integration. Participants were able to chain together multiple actions to locate interesting content as the recommender surfaced interesting possibilities. However, one problem with voice input to a recommender is that users do not necessarily know what they can ask for. Future work should investigate the ways of suggesting voice commands for finding recommendations, possibly in a personalized way~\cite{wiebe_exploring_2016, li_design_2011}.

Participants in our study encountered failures in voice recognition that highlight the challenges of specialized vocabulary in the movie domain. Although speech-to-text has become amazingly accurate overall~\cite{xiong_microsoft_2017}, we observed challenges around multilingual (``the movie `Coco’'' $\rightarrow$ ``the movie `Cocoa'''), proper noun (``Ralph Fiennes'' $\rightarrow$ ``Ray Fines''), and non-word searches (``The Intouchables'' $\rightarrow$ ``The Untouchables''). Recent work~\cite{springer_play_2018} developed an innovative method for detecting voice-inaccessible music, and a similar approach could be used in this domain (assuming the existence of massive search logs). Some aspects of this problem (e.g., selecting one movie out of a list) can be solved more simply by annotating the screen with easy-to-say labels; current examples of this pattern may be seen on voice-controlled TVs~\cite{haselton_amazon_2018} and mobile phones~\cite{corbett_what_2016}. 
These challenges suggest future work in domain-specialized speech-to-text models, and also in interface designs that expect failures and allow for more graceful patterns of correction.

Finally, our system highlighted the importance of familiarity in navigating recommendations. While many recommender systems are explicitly designed to filter out familiar content, we found that the presence of familiar movies reinforced participants' perceptions of recommendation quality and encouraged deeper exploration. 
The information gap theory of curiosity~\cite{loewenstein_psychology_1994} provides one theoretical explanation: the mixture of familiar and unfamiliar items causes the user to want to learn more about the unfamiliar. Future work should consider how to find an optimal balance of familiarity and novelty for each user~\cite{kapoor_i_2015} to simultaneously facilitate exploration and content-finding.

\section{Conclusion}
While voice user interfaces for TVs are becoming widely available, current systems treat voice input as an optional, secondary input, and rarely include voice output. To learn about the opportunities of voice-controlled TVs, we built a prototype system where the only input modality is voice. We conducted interview sessions to learn about how people might interact with such a system, and to develop a set of themes to guide future designs. We find that in a voice-only system, spoken inputs and outputs can be slow or repetitive, and that a key design guideline is to streamline these interactions. However, we also see a design tension between the competing needs for efficiency and expressiveness, pointing to future work in systems that can detect and adapt their output to user preferences and context. Finally, we see promise in supporting natural language interactions with recommendation technology; our design research underscores the importance of displaying the right balance of familiar and novel content, and suggests a new research direction around context- and user-personalized voice command suggestions.

\section{Acknowledgments}

This material is based on work supported by a grant from Amazon, the National Science Foundation under grant IIS-1319382, and
by the University of Minnesota’s Undergraduate Research Opportunities Program.


\bibliographystyle{SIGCHI-Reference-Format}
\bibliography{movie-tv-paper} 

\end{document}